# Ultrafast THz Field Control of Electronic and Structural Interactions in Vanadium Dioxide

A. X. Gray[1,2,*], M. C. Hoffmann[3], J. Jeong[4], N. P. Aetukuri[4], D. Zhu[3], H. Y. Hwang[5], N. C. Brandt[5], H. Wen[6], A. J. Sternbach[7,8], S. Bonetti[1], A. H. Reid[1], R. Kukreja[1,9], C. Graves[1,10], T. Wang[1,9], P. Granitzka[1,11], Z. Chen[1,12], D. J. Higley[1,10], T. Chase[1,10], E. Jal[1], E. Abreu[7,13], M. K. Liu[7,14], T.-C. Weng[15], D. Sokaras[15], D. Nordlund[15], M. Chollet[3], H. Lemke[3], J. Glownia[3], M. Trigo[1], Y. Zhu[6], H. Ohldag[15], J. W. Freeland[6], M. G. Samant[4], J. Berakdar[16], R. D. Averitt[7,8], K. A. Nelson[5], S. S. P. Parkin[4], H. A. Dürr[1,*]

**Vanadium dioxide, an archetypal correlated-electron material, undergoes an insulator-metal transition near room temperature that exhibits electron-correlation-driven and structurally-driven physics. Using ultrafast optical spectroscopy and x-ray scattering we show that these processes can be disentangled in the time domain. Specifically, following intense sub-picosecond electric-field excitation, a partial collapse of the insulating gap occurs within the first ps. Subsequently, this electronic reconfiguration initiates a change in lattice symmetry taking place on a slower timescale. We identify the kinetic energy increase of electrons tunneling in the strong electric field as the driving force, illustrating a novel method to control electronic interactions in correlated materials on an ultrafast timescale.**

Ultrafast control of fundamental electronic and structural interactions in strongly-correlated oxide materials is a promising avenue towards realizing the next generation of faster and more energy-efficient electronic devices [1-5]. Vanadium dioxide ($VO_2$) is viewed as a potential key building block for such devices due to its near-room-temperature insulator-metal transition (IMT), which can be triggered on a sub-picosecond time scale [6-8]. Such rapid switching, along with dramatic changes in the electrical and optical properties [9-11], gives rise to a multitude of novel possibilities for logic and memory devices, some of which could potentially be transformative to modern technology and computing [2,12,13].

While a full understanding of the IMT in $VO_2$ remains a challenge, it is clear that both electron-electron correlations and electron-lattice interactions are relevant [14,15]. What is less clear is the extent to which correlation and structure can be disentangled, though evidence is mounting that a monoclinic metallic phase can be obtained under appropriate conditions [16-20]. As such, a concerted effort aimed at understanding and achieving decoupling of the electronic and structural transitions in $VO_2$ has gained significant momentum [17,18,21]. To further investigate these intriguing possibilities, it is crucial to interrogate the dynamics of the electronic and lattice structure under well-controlled excitation conditions.

Along these lines, recent developments in the field of ultrashort high-field THz pulse generation opened the door for suppressing structural transformations in insulating $VO_2$ by nonresonant THz excitation of the electronic system at photon energies far below any of the relevant optical phonon modes or interband transitions [8,20]. A recent THz-pump THz-probe time-resolved study revealed that the IMT in $VO_2$ can, in fact, be initiated with a THz electric field [22]. This experiment revealed that the transition is a multi-step process. However, the sub-picosecond dynamics could not be temporally resolved nor were the structural dynamics measured.

We use THz excitation to initiate the IMT, which is probed using near-IR spectroscopy in combination with ultrafast hard x-ray scattering at the Linac Coherent Light Source (LCLS) [23,24]. We demonstrate that the electric-field-induced electronic and structural phase transitions in $VO_2$ can occur on different time scales. Electronic structure switching in a $VO_2$(001) film is observed virtually simultaneously with the sub-picosecond metamaterial-enhanced THz pulse, and is induced by electric-field-assisted tunneling of valence electrons into the conduction band. For sufficiently high electric fields $E \geq E_{threshold}$ (~1 MV/cm), this triggers the onset of the full transition to the rutile metallic state.

In stark contrast with the ultrafast electronic response, the onset of the change in lattice symmetry driven by the THz electric-field near this threshold value (1 MV/cm) is observed after a 1-3 ps delay, commencing with the dilation of vanadium dimers. Thus, during the first 1-3 ps after THz excitation a purely electronic fingerprint of an IMT in $VO_2$ is observed in the monoclinic phase, without detectable changes in lattice symmetry. These observations suggest that purely electronic switching of conductivity in strongly-correlated oxides may be possible without energy-dissipative lattice transformations, which has a far-reaching impact on future energy-efficient electronic devices utilizing ultrafast electronic switching.

For our first experiment, a high-quality polycrystalline 200 nm $VO_2$(001) thin film was grown epitaxially on an $Al_2O_3$(10$\bar{1}$0) substrate by pulsed laser deposition [25]. The resulting bulk-like $VO_2$ film was characterized using static temperature-dependent electronic transport and x-ray diffraction measurements revealing the expected IMT and monoclinic-rutile structural transformation at 337 K [26].

Ultrafast THz-pump near-IR- and hard x-ray probe measurements were carried out non-concomitantly in the experimental geometries shown in Fig. 1A. Single-cycle THz pump pulses with peak *E*-field of 0.4 MV/cm and central frequency of 0.6 THz were generated using the 20mJ output of a 1 kHz, 100 fs Ti:Sapphire amplifier via optical rectification in a $LiNbO_3$ crystal with the tilted pulse-front technique [27-29]. Metamaterial antennae consisting of an array of 8.5 μm-wide Au lines (Fig. 1A) were deposited on the samples via optical lithography in order to enhance the THz $E$-field at the sample. The antenna geometry was specifically optimized for the LCLS experiment, in order to provide 4× enhancement of the THz $E$-field (nominally 0.4 MV/cm) inside the 1.5 μm antenna gap [26]. Thus, maximum $E$-field peak strengths of up to 1.6 MV/cm (sufficient to initiate the IMT [22]) were achieved inside the gap in the normal-incidence geometry.

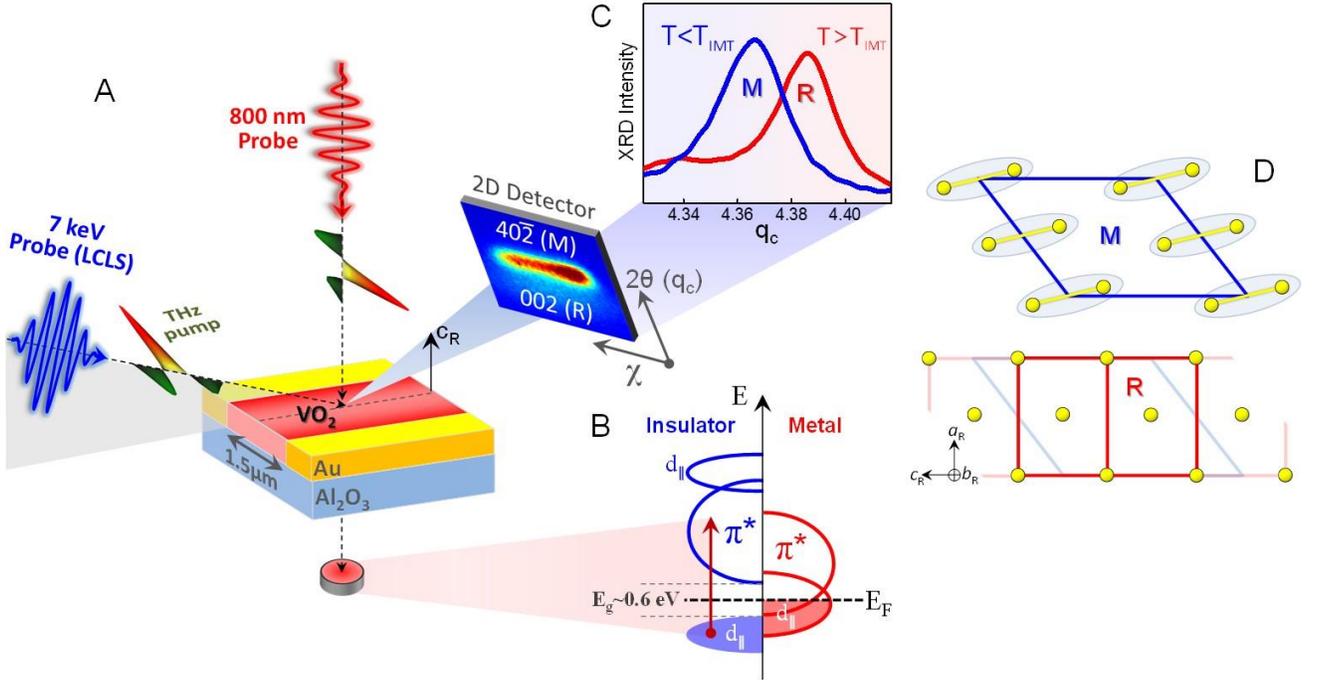

**Fig. 1.** Schematic of the THz-pump x-ray probe experiment at LCLS and relevant observable processes. (**A**) The single-cycle THz pump pulse and the 7 keV hard x-ray probe pulse are incident on the sample in collinear Bragg-diffraction geometry, while the THz-pump 800-nm probe measurements are carried out in a normal-incidence transmission geometry. The electric field of the THz pulse is locally enhanced at the sample surface using linear arrays of plasmonic metamaterial antennae [26]. (**B**) Schematic diagram of the equilibrium $VO_2$ band structure near the Fermi level in its insulating and metallic states. (**C**) Static θ-2θ spectroscopic calibration measurements of the monoclinic ($40\bar{2}$) and rutile (002) diffraction peaks (marked M and R, respectively) at temperatures well below and above $T_{IMT}$ using the 2D detector. (**D**) In the low-temperature (T<$T_{IMT}$) phase $VO_2$ has a $P2_1/c$ monoclinic crystal structure characterized by dimerization of the neighboring V atoms along the $c_R$ direction. In the high-temperature (T>$T_{IMT}$) phase the lattice undergoes a structural transition to a higher-symmetry $P4_2/mnm$ rutile structure, clearly discernable via hard x-ray scattering (see **A** and **C**).

In order to measure the time-resolved electronic response of $VO_2$, THz-pump near-IR-probe measurements were carried out in collinear normal-incidence geometry as shown in Fig. 1A. Part of the 800 nm beam generated by the Ti:Sapphire amplifier was utilized as a probe of transmission with a standard ultrafast InGaAs diode used as a detector. The wavelength of 800 nm is sensitive to the interband transitions from the occupied $d_\parallel$ ($a_{1g}$) orbitals across the insulating gap and into the unoccupied $\pi^*$ ($e_g^\pi$) states (see Fig. 1B). During the IMT, bonding and antibonding $d_\parallel$ states merge into a single band, while the $\pi^*$ band shifts to lower energies. The two bands overlap in energy, and the resulting non-zero density of states at the Fermi level accounts for the metallic behavior and change in the transmission signal at 800 nm [15].

As shown in Fig. 2, an electronic response is observed virtually simultaneously with the 0.5 ps-long (FWHM) THz excitation pulse and exhibits a rich multi-step evolution. In order to decouple the distinct electronic processes taking place within the first several ps following THz excitation, we carried out a series of temperature- and fluence-dependent THz-pump near-IR-probe measurements. Figures 2A and 2B depict time-delay traces of normalized near-IR transmission (at λ=800 nm) recorded as a function of temperature and THz pump excitation, which is expressed in terms of the peak $E$-field. The THz pulse waveform (not enhanced) was measured and calibrated via standard electro-optic sampling [30], as shown in the outset plot above Fig. 2A, and has a typical profile observed in prior studies [22,31,32]. The maximum achieved peak $E$-field of 1.6 MV/cm was normalized to 100%, and the attenuated peak $E$-fields of 80% (1.28 MV/cm), 60% (0.96 MV/cm), *etc.* with respect to this nominal value were obtained using a pair of rotatable THz wire-grid polarizers.

Figure 2A shows $E$-field-dependent pump-probe time-delay traces recorded at room temperature (300 K), which is 37K below the measured $T_{IMT}$ of 337 K. As the data shows, at temperatures far below the onset of the IMT, the dynamics are
2

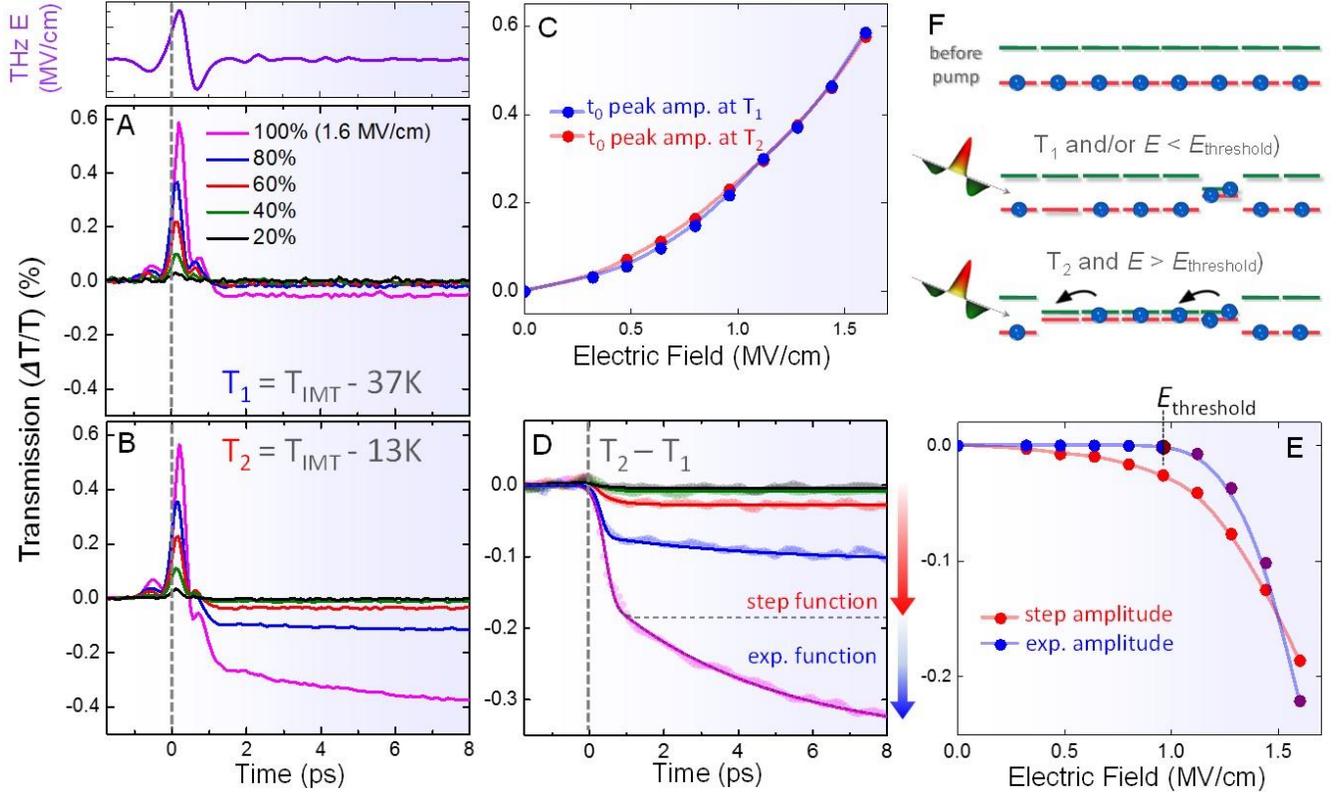

**Fig. 2.** Electronic dynamics in a $VO_2$ thin film undergoing an IMT induced by an ultrafast THz electric-field pulse (shown in the outset). (**A**) The *E*-field dependent electronic response measured via near-IR transmission at room temperature ($T_1$=300K) is dominated by a sharp transient increase in transmission consistent with Wannier-Stark-like band renormalization [3,33-38]. For high fluences (*e.g.* purple delay trace) this process is accompanied by a weak concomitant drop in transmission, characteristic of sudden metallization of $VO_2$ via direct Zener-Keldysh-like interband tunneling [38-42]. (**B**) At the onset of the IMT ($T_2=T_{IMT}$-13 K) tunneling processes become more efficient, and above a threshold peak-field of approximately 1 MV/cm the absorbed energy density is sufficient to initiate lattice dynamics (see Fig. 3 and [22]). (**C**) Wannier-Stark peak amplitude exhibits a non-linear dependence on the peak E-field with minimal temperature dependence between 300K and 324K. (**D**) The subsequent dynamics can be pictured more clearly by subtracting off the temperature independent dynamics. (**E**) *E*-field dependence of the amplitudes characterizing the pulse-width-broadened step and exponential functions suggests different electronic origins of the two processes and reveals the threshold onset of the exponential feature. (**F**) Schematic representation of the early time electronic dynamics (see text for details).

dominated by processes confined within the temporal window during which the THz pulse is present in the sample (~1 ps). There is a substantial increase in the optical transmission of the film that follows in shape the square of the THz pulse waveform. This suggests an electric field-driven modification of the electronic structure leading to a transient decrease of the near-IR absorption analogous to what has been observed in dielectrics by Schultze *et al.* [3]. Essentially, Wannier-Stark-like renormalization of the band structure results in a decrease in the joint density of states between occupied and unoccupied levels leading to the observed transmission increase [33-38]. The on-off behavior, as well as the nonlinear dependence on the strength of the *E*-field (Fig. 2C) is consistent with prior experimental observations as well as with the theoretical picture [3,38]. Furthermore, at strong enough fields, Wannier-Stark-like renormalization results in the mixing of the valence- and conduction-band electronic states leading to Zener-Keldysh-like interband tunneling of the valence electrons at the valence-conduction-band anticrossings [38-42].

We observe an experimental fingerprint of such non-thermal electronic tunneling process, characterized by an abrupt decrease in transmission due to prompt metallization coinciding with the transient peak attributed to the band renormalization. We emphasize that this effect is simultaneous with the exciting THz pulse, and ceases once the THz pulse is no longer present in the sample (past *t*=1 ps). Importantly, this step-like tunneling feature is (*unlike* the *E*-field dependent band renormalization) highly temperature-dependent as evidenced by the differences between the data at $T_1$=300 K (Fig. 2A) and $T_2$=324 K (Fig. 2B). We speculate that the apparent increase in the efficiency of the interband tunneling processes near the $T_{IMT}$ (Fig. 2B) could be attributed to changes in the electronic structure and, in particular, to the softening of strong electronic correlations at the onset of the IMT observed recently via temperature-dependent polarized x-ray absorption spectroscopy [43,44].

At room temperature (300 K), the interband tunneling at high THz peak-fields (*e.g.* 1.6 MV/cm) induces a stable long-lived conducting state that persists for picoseconds after the THz pulse (see purple delay trace in Fig. 2A). This observation is consistent with the recent work by Mayer *et al.* [45] wherein the IMT in $VO_2$ was induced by multi-THz transients. As we tune the temperature closer to $T_{IMT}$, tunneling becomes, as mentioned above, more efficient (see respective delay traces in Fig. 2B). Finally, above a certain THz peak electric field between 0.9 and 1.1 MV/cm a *third* process sets in, characterized by an exponential decrease in near-IR transmission after $t$=1 ps. Above this threshold, this exponential process dominates the IMT dynamics, driving the film towards the rutile metallic phase over tens of picoseconds [26]. Observation of these slower dynamics is consistent with the measurements by Liu *et al.* [22] (also at the temperature of 324 K) where the energy density deposited by the THz pulse is sufficient to drive electron-phonon thermalization and heating towards (and above) $T_{IMT}$. We consider this picture in more detail in the following paragraphs.

Since the Wannier-Stark-like renormalization does not show a significant temperature-dependence, as evidenced by the overlaid plots of the *E*-field-dependent peak intensities for 300 K and 324 K in Fig. 2C, we can subtract this contribution from the higher-temperature data using respective room-temperature spectra. This data analysis reveals the form and amplitude of the tunneling breakdown dynamics in $VO_2$ during the first picosecond of the IMT (Fig. 2D). The abrupt decrease in transmission follows a simple step function broadened by the THz pulse duration, consistent with ultrafast non-thermal field-induced tunneling as described above [45]. For high THz fluences this process is followed by a slower metallization dynamics characterized by an exponential decay (1-8 ps in Fig. 2D).

Fig. 2E plots the amplitude versus field-strength of the step-like and exponential features from Fig. 2D. This different functional dependence of these two features suggests different electronic origins. The step-function amplitude (red curve) exhibits exponential behavior (versus field) consistent with the nonresonant Zener tunneling equation [39,41] and is measurable even at moderate peak-fields (0.32 MV/cm). In contrast to this continuous field-dependence, a clear threshold behavior is observed for the exponent amplitude (blue curve), suggesting presence of a competing mechanism preventing full metallization of $VO_2$ following THz excitation with fields below the threshold. The threshold field sufficient to drive the electronic system through the point of no return is estimated to be approximately 1 MV/cm (blue/purple marker). Above this critical field, the IMT progresses exponentially with steep field-dependence (purple markers) resulting in metallization of a significant fraction of the film.

The results of our THz-pump visible-probe experiments suggest the following scenario (Fig. 2F): Below $E_{threshold}$ (and above ~0.32MV/cm), the applied field induces Wannier-Stark band renormalization accompanied by Zener-Keldysh-like interband tunneling. At times longer than the pulse duration, the flat plateau that is observed arises from carriers that have tunneled and not recombined. In the language of correlated electron materials, there is double occupancy with a finite density of states at $E_F$ – that is, there is a partial collapse of the Mott-Hubbard gap. In this lower *E*-field range and on the timescales we have measured there is no observable increase or decrease in the signal beyond 1 ps. As we raise the sample temperature closer to the onset of the IMT (to $T_2=T_{IMT}-13K$) electronic correlations [43] are softened, which leads to more efficient tunneling and to a commensurate increase in the carrier density. The THz electric field accelerates these carriers, which subsequently relax though electron-phonon collisions. For sufficiently high fields, above $E_{threshold}$, this increases the lattice temperature initiating the IMT towards the rutile metallic phase. This corresponds to the exponentially decreasing term in Fig. 2D (magenta line).

This picture suggests that the monoclinic phase can be metallized without a structural change for fields below $E_{threshold}$, while for fields above $E_{threshold}$ a monoclinic metallic phase is a dynamic precursor to the rutile metallic state. To verify that the structural dynamics are in accordance with this interpretation, we have utilized THz-pump hard x-ray probe scattering experiments near the threshold THz fields, as described in the following paragraphs.

In order to measure the time-resolved lattice response of $VO_2$, the THz pump was incident on the sample collinearly with the 7 keV FEL x-ray beam at an incidence angle of 38º, which is the Bragg angle for the rutile (002) diffraction peak, sensitive to the changes in atomic-layer spacing along the $c_R$ axis. In the static case, a temperature-induced transition results in a decay of the rutile (002) peak (R) and growth of the monoclinic ($40\bar{2}$) peak (M), which is offset in *q*-space by 0.02 nm$^{-1}$ due to the change in the average lattice spacing (see Fig. 1C). A large two-dimensional position-sensitive detector (CSPAD) [46] enables θ-2θ spectroscopic measurements of the rutile and monoclinic diffraction peaks by rotating the sample (θ) and measuring the intensities of the specular Bragg reflection spots (R and M) while tracking their movement across the detector (2θ). The structural transition accompanied by dimerization of the neighboring V atoms along the out-of-plane $c_R$ direction and tilting of the resultant V-V dimers along the rutile [110] and [$1\bar{1}0$] directions is depicted schematically in Fig. 1D.

Figure 3 shows the temporal evolution of the monoclinic-rutile structural transformation during the first 500 ps following THz pump excitation. Data points were obtained by integrating the intensities of the monoclinic and rutile scattering peaks measured via θ-2θ scans at a given time delay. The measurements were carried out at the sample temperature of 324 K (T$_{IMT}$-13 K), at which approximately 2.7% of the VO$_2$ film under the x-ray measurement spot has already undergone the transition [26]. Thus, prior to the THz electric-field pulse arrival at t = 0 ps, approximately 97.3% of the sample underneath the x-ray probe spot was still in the low-temperature monoclinic phase, and approximately 2.7% of the sample was already in the high-temperature rutile phase.

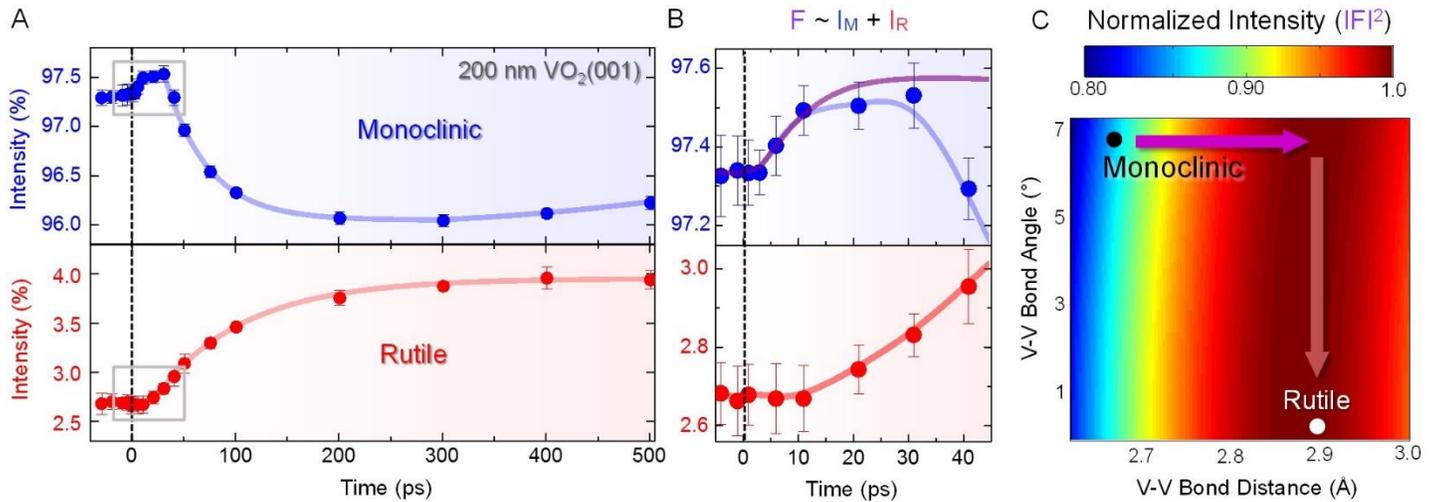

**Fig. 3.** Structural dynamics of the polycrystalline VO$_2$ (001) thin film undergoing an IMT induced by an ultrafast THz electric-field pulse. (**A**) Long-term temporal evolution of the monoclinic and rutile peak intensities, obtained via θ-2θ scans at each time delay. (**B**) Measurements in the vicinity of time-zero suggest no detectable change in the lattice symmetry (from monoclinic to rutile) during the first few ps following the THz pulse. The first 30 ps of the structural transition are characterized by the emergence of a transient monoclinic phase with a higher structure factor F, followed by the decay of the monoclinic peak and the concomitant growth of the rutile peak. (**C**) Structure factor calculations suggest a two-step M-R transition pathway proposed by Baum *et al.* [47], commencing with the dilation of the V-V dimers (purple horizontal arrow) and followed by a change in the V-V bond angle resulting in the final rutile symmetry (red vertical arrow).

In the grazing-incidence Bragg geometry necessary for this measurement (Fig. 1A), the *E*-field peak strength is estimated to be approximately 1.0 MV/cm using basic geometric considerations [26]. At this field strength, the THz excitation drives VO$_2$ towards the threshold above which electron-lattice interactions trigger the change in lattice symmetry (Figs. 2D-E). Thus, the long-term (hundreds-of-picoseconds) structural dynamics shown in Fig. 3A are characterized, as expected, by the decay of the monoclinic phase and the concomitant growth of the rutile phase. However, of particular interest are the first 45 ps of the THz-driven transition, shown in Fig. 3B. The change in lattice symmetry (from monoclinic to rutile) is virtually undetectable during the initial 3 ps following the THz pump pulse. Subsequently, the onset of the lattice transformation occurs, and manifests, initially, as an increase of the monoclinic peak intensity. The increase of the monoclinic peak intensity continues for 30 ps before eventually decreasing and, further, is not accompanied by a decrease in the intensity of the rutile peak. This suggests that a new monoclinic phase of VO$_2$, characterized by a higher structure factor (F), emerges at the onset of the structural transition. In fact, the sum of the intensities of the monoclinic and rutile peaks, which is directly proportional to F (Fig. 3B), grows during the first 30 ps of the structural transition and then remains constant as the increase of the intensity of the rutile peaks begins to compensate for the decay of the intensity of the monoclinic peak (after $t \approx 30$ ps).

The multi-step dynamics in the structural evolution of the dimerized monoclinic VO$_2$ lattice undergoing the IMT can be readily explained using the two-step pathway proposed by Baum *et al.* [47] which can be mapped onto the 2D map of normalized intensity $|F|^2$ plotted as a function of the V-V bond distance and angle (Fig. 3C). The electron diffraction data [45] suggests that the optically-induced transition from the monoclinic to the rutile structural phase commences with a dilation of the V-V dimers along the direction of the chemical bonds, which results in the initial increase of the structure factor (horizontal arrow in Fig. 3C), followed by a change in the V-V bond angle that leads to the final rutile symmetry. Our results confirm this scenario for both the near-IR [26] and THz excitation.

Most importantly, the combined results of our THz-pump near-IR and x-ray probe studies demonstrate that the electronic and structural transitions in VO$_2$ can be decoupled in the time domain, and that for several picoseconds following the THz excitation a metallic phase of VO$_2$ can be supported by a monoclinic lattice, consistent with several recent static and time-resolved studies [17,18,48,49].

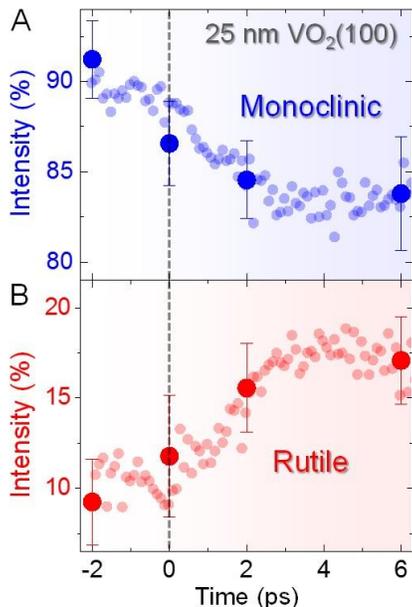

**Fig. 4.** Structural dynamics of a thin epitaxially-strained 25 nm VO$_2$(100) film undergoing an IMT induced by an ultrafast THz electric-field pulse shows a more efficient way of coupling THz excitation to the lattice. (**A**) Ultrafast evolution of the monoclinic (040) peak (insensitive to dimer dilation) measured in the same experimental geometry but at the photon energy of 9 keV shows a ps-scale response to the THz excitation. Concomitant growth of the rutile (400) peak is shown in (**B**). Measurements are obtained in two modes: θ-2θ scans (large blue/red circles) recorded at each time delay, as well as continuous delay-trace intensity measurements at the maxima of the (040)$_M$ and (400)$_R$ peaks (light blue/red dots).

Time-resolved dynamics of the rutile Bragg peak width suggests that the conducting rutile phase starts emerging from the monoclinic matrix in the form of nm-scale islands [26]. Such islands have been previously observed statically during temperature-driven transition by scanning near-field infrared microscopy [50] and scanning probe x-ray diffraction [51].

Further investigations of the sub-picosecond dynamics near time-zero are required to interrogate the possible motion of the monoclinic lattice during the intense THz pulse, which reaches its peak value at $t$=+0.2 ps (within a 2 ps-long window between the measured time delays of -1 ps and +1 ps). A high-statistics delay-trace of the rutile peak intensity [26], however, suggests that no detectable changes in lattice symmetry occur during this time interval, at least in the rutile phase. Finally, effects of the multiple THz reflections from the back surface of the substrate need to be considered in future studies.

Along these lines, we discover that the temporal dynamics of the THz-driven structural transition in VO$_2$ is highly-tunable via thickness and choice of substrate, which determine the crystallographic orientation and strain of the film. A second set of ultrafast THz-pump hard-x-ray diffraction probe measurements carried out on a thin epitaxially-strained 25 nm VO$_2$(100) film deposited on TiO$_2$(100) substrate reveal more efficient coupling of THz excitation to the lattice, resulting in a nearly-instantaneous structural response (Fig. 4).

Although for this crystallographic orientation a measurement of the (400)$_R$ peak is not sensitive to the dimer dilation, we can nevertheless track the temporal evolution of the M and R phases following the THz pump by tracking the intensities of the corresponding (040)$_M$ and (400)$_R$ peaks, which are sensitive to changes in atomic-layer spacing along the $a_R$ axis, perpendicular to the axis of dimerization ($c_R$).

Results of the time-resolved THz-pump x-ray diffraction probe measurements carried out at the temperature of 334 K (10 K below the transition temperature for the VO$_2$(100) film [26]) are shown in Figs. 4A and 4B. At this temperature, approximately 10.5% of the film under the x-ray spot has already undergone the M-R structural transition, as evidenced by the static x-ray diffraction measurements carried out immediately before the time-resolved pump-probe delay scans [26]. Ultrafast response and concomitant decay (M) and growth (R) of the monoclinic and the already-nucleated rutile phases is shown in Figs. 4A and 4B respectively. In these plots we combine the time-resolved M and R peak intensities obtained via θ-2θ scans (large blue/red circles) with the finer-time-step normalized delay traces obtained simply by sitting on the peak maxima and recording time-dependent count rates (light blue/red dots). Such a comparison allows us to discern ps-scale structural dynamics near time-zero and lessens the problem of low count rates due to a much smaller film thickness.

In contrast with the thicker (200 nm) polycrystalline VO$_2$(001) film deposited on Al$_2$O$_3$(10$\bar{1}$0), a structural response of the thinner (25 nm) epitaxial VO$_2$(100) film to the *E*-field excitation occurs virtually simultaneously with the THz pulse (Fig. 4). The M-R transition progresses monotonically during and after the pulse, with approximately 7.5% of the film volume being converted into rutile phase during the first 4 ps. The higher initial temperature is likely to be a contributing factor to the faster onset and increased speed of the structural response, when comparing the delay traces in Figs. 3 and 4 for the (001) and (100) films respectively. Such contrasting structural dynamics is furthermore not surprising and perhaps even expected considering differences in strain [11], thickness [52], substrate [53], and different phase separation scenarios for the two crystalline orientations [48,50,54]. While further studies are required to ascertain the origin of this rapid structural change, these results nevertheless reveal that epitaxial tuning of the initial state strongly affects the dynamics.

In conclusion, our experiments show that non-equilibrium electronic and structural dynamics in VO$_2$ can be disentangled in the time domain by inducing the IMT with an intense non-resonant THz pulse and probing with near-IR and hard x-ray pulses. The primary trigger for the electric-field-driven transition is shown to be Zener-Keldysh tunneling assisted by the Wannier-Stark band renormalization. At intermediate fields a conducting monoclinic phase is established. At the temperatures sufficiently close to T$_{IMT}$ and for THz peak-fields above a threshold of approximately 1 MV/cm the dynamically established precursor monoclinic metallic phase evolves towards the conventional rutile metallic phase. This is a result of softening of electronic correlations and of efficient energy transfer to the lattice from carrier acceleration and electron-phonon collisions. The rich multi-dimensional landscape resulting from the interplay of the Mott and Peierls physics provides a pathway to achieving purely-electronic THz switching in strongly-correlated oxides via ultrafast electric-field excitation.

**ACKNOWLEDGEMENTS**

Research at Stanford was supported through the Stanford Institute for Materials and Energy Sciences (SIMES) under contract DE-AC02-76SF00515. Use of the Linac Coherent Light Source (LCLS), SLAC National Accelerator Laboratory, is supported by the U.S. Department of Energy, Office of Science, Office of Basic Energy Sciences under Contract No. DE-AC02-76SF00515. Research by the MIT group was supported in part by Office of Naval Research Grant No. N00014-13-1-0509 and National Science Foundation Grant No. CHE-1111557. Work at Argonne was supported by the US Department of Energy, Office of Science, Office of Basic Energy Sciences, under Contract No. DE-AC02-06CH11357. R.D.A. acknowledges support from the U.S. Department of Energy, Office of Science, Office of Basic Energy Sciences under Grant No. DE-FG02-09ER46643. E.A. acknowledges support from Fundação para a Ciência e a Tecnologia, Portugal, through doctoral degree fellowship SFRH/ BD/ 47847/ 2008. J.B. is supported by the DFG under grant SFB762. S.B. acknowledges support from the Knut and Alice Wallenberg Foundation.

**AUTHORS AFFILIATIONS**

[1]Stanford Institute for Materials and Energy Sciences, SLAC National Accelerator Laboratory, Menlo Park, California 94025, USA. [2]Department of Physics, Temple University, Philadelphia, Pennsylvania 19130, USA. [3]Linac Coherent Light Source, SLAC National Accelerator Laboratory, Menlo Park, California 94025, USA. [4]IBM Almaden Research Center, San Jose, CA 95120, USA. [5]Department of Chemistry, Massachusetts Institute of Technology, Cambridge, Massachusetts 02139, USA. [6]Advanced Photon Source, Argonne National Laboratory, Argonne, Illinois 60439, USA. [7]Department of Physics, Boston University, Boston, Massachusetts 02215, USA. [8]Department of Physics, The University of California at San Diego, La Jolla, California 92093, USA. [9]Department of Materials Science and Engineering, Stanford University, Stanford, California 94305, USA. [10]Department of Applied Physics, Stanford University, Stanford, California 94305. [11]Van der Waals-Zeeman Institute, University of Amsterdam, 1018XE Amsterdam, The Netherlands. [12]Department of Physics, Stanford University, Stanford, California 94305. [13]Institute for Quantum Electronics, ETH Zürich, 8006 Zürich, Switzerland. [14]Department of Physics and Astronomy, Stony Brook University, Stony Brook, NY 11794, USA. [15]Stanford Synchrotron Radiation Lightsource, SLAC National Accelerator Laboratory, Menlo Park, California 94025, USA. [16]Institut fur Physik, Martin-Luther-Universität Halle-Wittenberg, 06099 Halle/Saale, Germany.

*Corresponding authors. E-mail: axgray@temple.edu, hdurr@slac.stanford.edu